\documentclass[11pt]{article}
\usepackage{moriond,epsfig}

\def\be{\begin{equation}}
\def\ee{\end{equation}}
\def\bea{\begin{eqnarray}}
\def\eea{\end{eqnarray}}

\begin{document}

%\begin{flushright}
%{\large CDF/PUB/CDF/PUBLIC/7653}
%\end{flushright}

\vspace*{4cm}
\title{NEW DIFFRACTIVE RESULTS FROM THE TEVATRON}

\author{ MICHELE GALLINARO \footnote{Representing the CDF and D\O~collaborations.}}

\address{%(for the CDF and D0 collaborations)\\
Laboratory of Experimental High Energy Physics, \\ The Rockefeller University, \\ 
1230 York Avenue, New York, NY 10021, USA \\
}

\maketitle\abstracts{
Experimental results in diffractive processes are summarized and a few notable characteristics described in terms of Quantum Chromodynamics.
Exclusive dijet production is used to establish a benchmark for future experiments
in the quest for diffractive Higgs production at the Large Hadron Collider.
Using new data from the Tevatron and dedicated diffractive triggers, no excess over a smooth falling distribution for exclusive dijet events could be found.
Stringent upper limits on the exclusive dijet production cross section are presented.
The quark/gluon composition of dijet final states is used to provide additional hints on exclusive dijet production.}

\section{Diffraction and Quantum Chromodynamics}

In terms of Quantum Chromodynamics (QCD), the exchange of no quantum numbers in hadronic interactions can be described by
the exchange of an ``object'', called the pomeron, interpreted as an entity composed of quarks and gluons, 
with the quantum numbers of the vacuum.
The theory of QCD has proved successful in explaining hard processes, such as those where
high transverse momentum jets are produced in the final state. These processes can be well described within the framework of perturbative effects.
In contrast, soft processes, such as elastic scattering and low-$p_T$ diffractive dissociation, are dominated by non-perturbative components, 
and probe the large distance behavior of quark and gluon interaction.
The treatment of low $Q^2$ processes is complicated as the strong coupling constant $\alpha (Q^2)$ grows at small $Q^2$
causing confinement~\cite{collins}.
Hard diffractive interactions, which contain both hard and soft processes, can provide
an important step in the understanding of the transition from
non-perturbative to perturbative QCD.
Inclusion of the pomeron in the framework of QCD is therefore a necessary step in the understanding of diffractive 
phenomena and of QCD itself, and attempts have been made in this direction~\cite{low,lipatov}.

Elastic and diffractive processes can be interpreted in terms of pomeron exchange and are characterized by a final
state where the surviving (leading) hadrons are emitted at very small angles with respect to the incoming beam, carrying a large
fraction of the initial momentum.
In this context, the Feynmann variable $x_F=p_L^\prime/p$ is a useful quantity in the discussion of inclusive hadronic 
interactions at large energies, where $p_L^\prime$ is the longitudinal momentum of the outgoing particle.
In a typical hadron scattering, most of the energy after the collision emerges through particles with a small deviation from 
the original direction of the incoming particle; longitudinal momenta can be large, 
while the average transversal component of the particles' momenta is relatively small ($p_T<1~\rm GeV/c^2$) and falls rapidly with 
increasing $p_T$.
Therefore, in the approximations in which $|p_T|\ll |p|\simeq \sqrt{s}/2$ and $|p_L|\simeq |p|$, 
one obtains $x_F\simeq 1- M^2/s$, where $M$ is the mass of the diffractive system.
Diffraction dissociation processes take place in a kinematic region of $x_F\simeq 1$.

Studies of diffractive events at the Tevatron~\cite{dino} have yielded a few salient characteristics in three main areas:
(i) the ratio of single diffractive (SD) to non-diffractive (ND) event rates, (ii) diffractive structure functions, and (iii) gap survival probability.

(i) Although the different processes studied in hard diffractive processes have different 
sensitivities to quarks and gluons in the pomeron, SD to ND ratios are found to be of similar magnitude ($\sim 1\%$) for all processes.
This suggests that the structure of the pomeron probed in SD events is similar to the structure of the proton probed in ND events.

(ii) Structure functions, i.e. the gluon and quark content of the interacting partons, 
can be investigated by comparing SD with ND events and are extracted from the ratio of SD to ND dijet production rates.
The diffractive structure function rises as the $x-$Bjorken scaling variable, $x_{\bar p}$~\cite{xbj}, decreases, and 
the normalization of the SD to ND ratio is suppressed by a factor of $\sim 10$ with respect to the expectation of the 
parton distribution functions obtained at HERA. This implies that the pomeron does not possess a 
universal process-independent structure function, but it must instead be convoluted with an additional effect, 
which spoils the formation of the rapidity gap.
The breakdown of factorization, as this effect is generally referred to, is not yet well understood theoretically.

(iii) The measurement of the two-gap production rate relative to that of the one-gap processes is not as
suppressed as the SD differential cross sections. 
This is valid for both soft and hard diffractive processes.
In fact, once the formation of the first gap is established, 
the second gap can be produced free of the color exchange constraints which spoil its formation.
The ratio of two-gap to one-gap event rates offers the possibility of studying the probability of forming a rapidity gap.
The gap formation probability is measured to be approximately 20\%, with a similar rate in soft and hard diffraction.
Furthermore, the measurement of DPE to SD event rates restores factorization for events that already have a rapidity gap.

\section{Exclusive Dijet Production}

Exclusive dijet production has been studied at the Tevatron during Run~II.
Double Pomeron Exchange (DPE) processes of this type have a clean experimental signature as
the final states are centrally produced with two large ($\Delta\eta>4$) rapidity gaps on opposite sides of the interaction region.
Thanks to their centrality, the final states are well measured.
First studied as a tool to characterize the nature of diffractive interactions, these events have also been used to measure the gap survival 
probability and more recently have been of interest to set a benchmark for rare exclusive processes, such as diffractive Higgs production.
The small production cross section of the latter can be inferred and predictions be set for future experiments at the Large Hadron Collider (LHC).

\begin{figure}[h]
\epsfxsize=1.0\textwidth
\centerline{
\psfig{figure=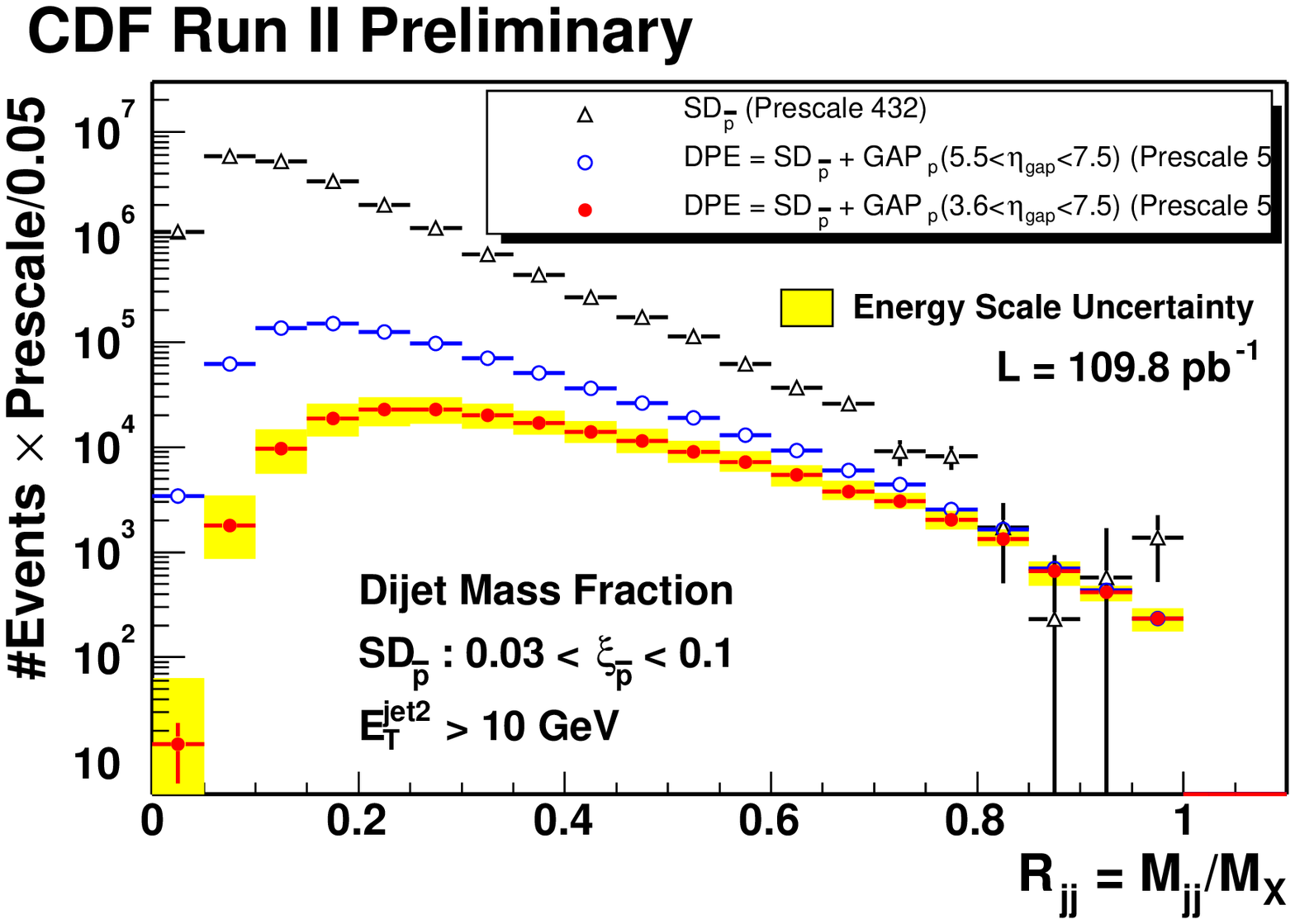,width=0.56\hsize}
\psfig{figure=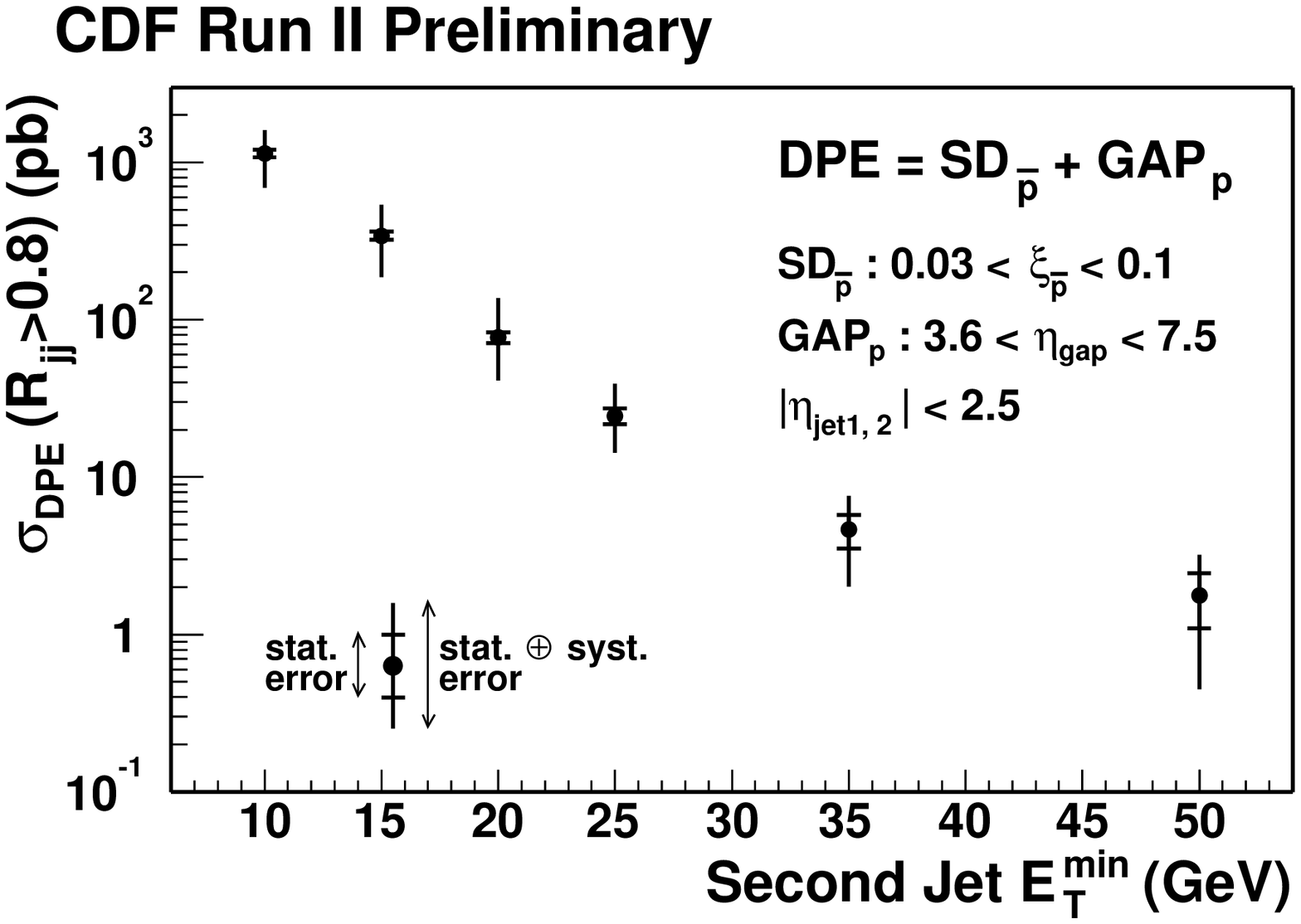,width=0.56\hsize}
}
\caption{\label{fig:dpe_massratio}
\small
{\em Left:} Dijet mass fraction for different rapidity gap regions.
{\em Right:} Upper limit cross section for DPE dijet events with $R_{jj}>0.8$ as a function of the minimum next-to-leading jet $E_T$.
}
\end{figure}

At CDF, in $\sim$110~pb$^{-1}$ of Run~II data, a triggered DPE dijet sample already consists of approximately 1.4 million events.
After selection cuts are applied to the event sample, a purer DPE sample contains approximately 34,000 events.
The dijet mass fraction ($R_{jj}$), defined as the dijet invariant mass ($M_{jj}$) divided by the mass of the entire system,
$M_X =\sqrt{\xi_{\overline{p}}\cdot\xi_p\cdot s}$, is calculated using all available energy in the calorimeter.
If jets are produced exclusively, $R_{jj}$ should be equal to one~\cite{nlo}.
Uncorrected energies are used in Figure~\ref{fig:dpe_massratio} (left) and no visible excess over a smooth distribution is evident 
at $R_{jj}\sim 1$.
After including systematic uncertainties, an upper limit on the exclusive dijet production cross section is calculated
based on all events with $R_{jj}>0.8$ (Table~\ref{tab:xs}). 
The measurement provides a generous upper limit cross section, as all events at $R_{jj}>0.8$ 
are considered to be due to exclusive dijet production.
The cross section upper limits on exclusive dijet production as a function of the minimum next-to-leading jet $E_T$ 
is presented in Figure~\ref{fig:dpe_massratio} (right).

\begin{table}[h]
\begin{center}
\begin{tabular}{c|c}
\hline
minimum leading jet $E_T$ & cross section upper limit \\ \hline
10 GeV & $1140\pm60({\rm stat})^{+470}_{-450}({\rm syst})$~pb \\
25 GeV & $ 25\pm3({\rm stat})^{+15}_{-10}({\rm syst})$~pb \\ \hline
\end{tabular}
\caption{\label{tab:xs} Exclusive dijet production cross section limit for events at $R_{jj}>0.8$.}
\end{center}
\end{table}

\section{Heavy Flavor Tagging}

\begin{figure}[h]
\epsfxsize=1.0\textwidth
\centerline{
\psfig{figure=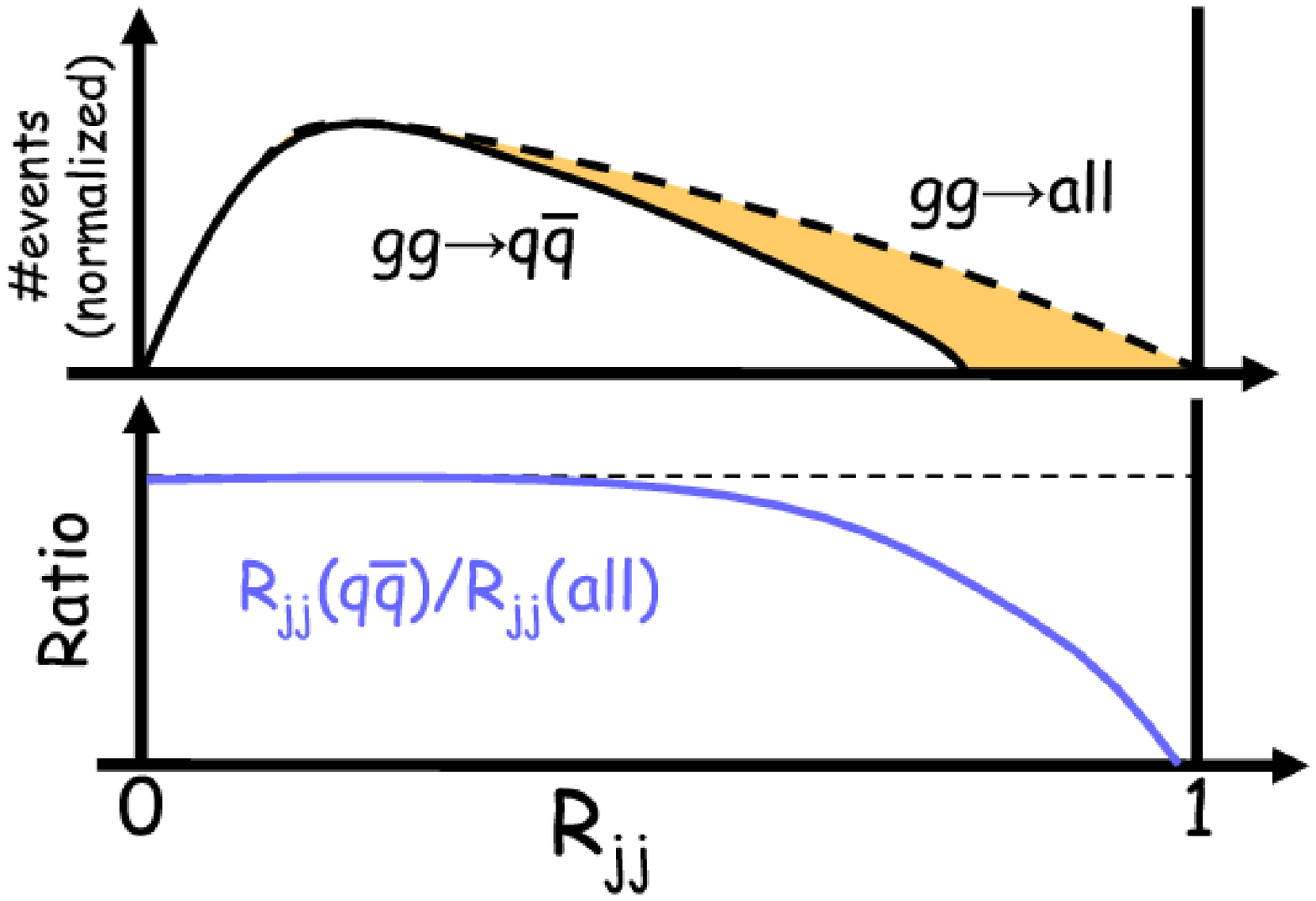,width=0.56\hsize}
\psfig{figure=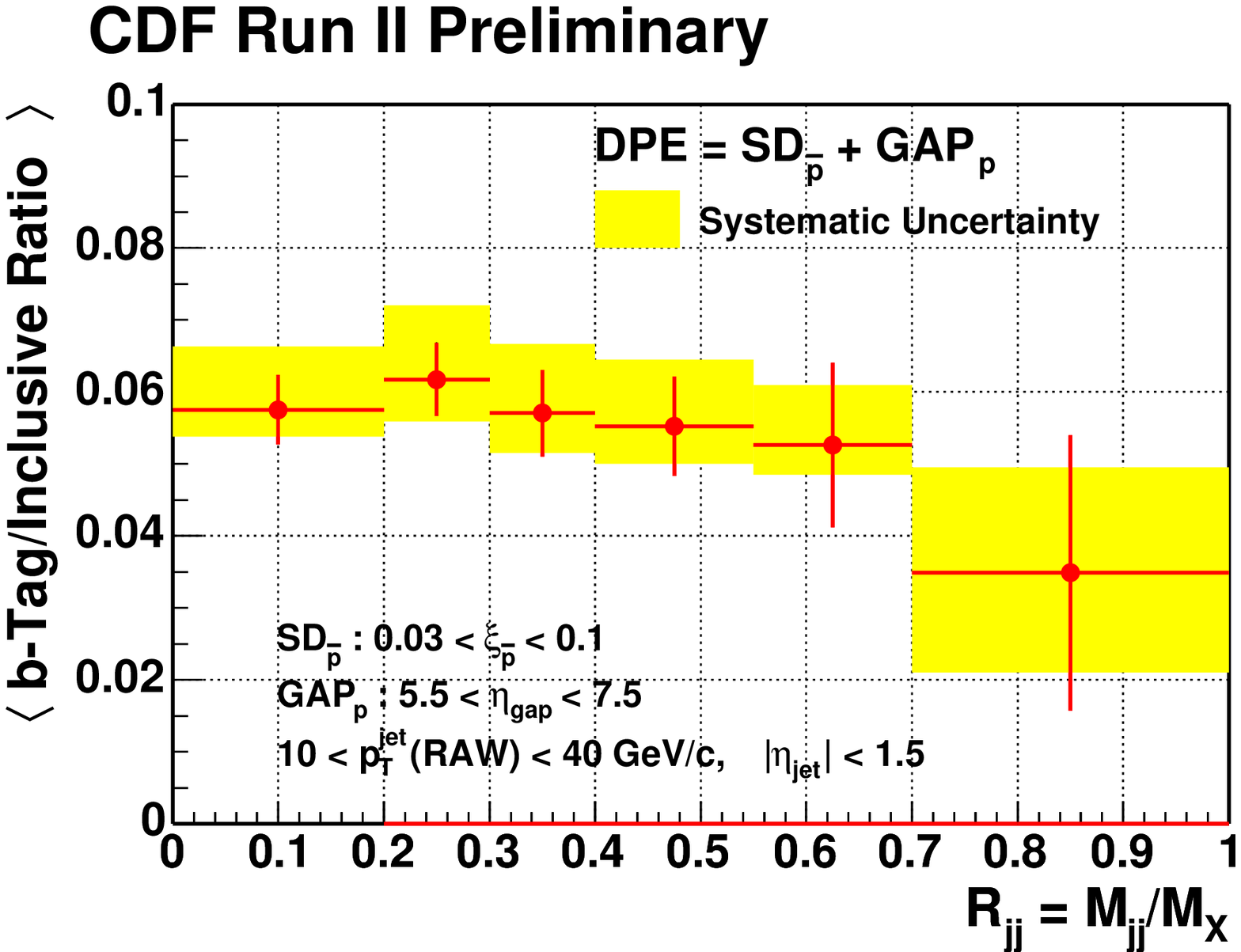,width=0.56\hsize}
}
\caption{\label{fig:exclusive_dpe}
\small
{\em Left:} Schematic illustration of the behavior of dijet events
processes as a function of the mass fraction, $R_{jj}=M_{jj}/M_X$, 
i.e. the ratio of dijet mass divided by the invariant mass of the entire system.
Top: 
The spectrum of $q\bar{q}$ events (solid line) is suppressed in the high (exclusive) 
$R_{jj}$ region with respect to all dijet events (dashed line).
Bottom: the ratio of $q\bar{q}$ to all dijets is expected to fall in the exclusive 
region ($R_{jj}\rightarrow 1$) if exclusive dijet events are produced.
{\em Right:} 
Ratio of b-tagged jets to all inclusive jets as a function of the mass fraction $R_{jj}$. The error bands correspond to the overall systematic uncertainty.}
\end{figure}

The quark/gluon composition of dijet final states can be used to provide additional hints on exclusive dijet production.
At leading order (LO), the exclusive $gg\rightarrow gg$ process is dominant, 
as the contribution from $gg\rightarrow q\bar{q}$ is strongly suppressed~\cite{khoze}.
In fact, the exclusive dijet cross section $\sigma_{excl} (gg\rightarrow q\bar{q})$ vanishes as $m(q)/p_T(q)\rightarrow 0$ ($J_z=0$ selection rule).
This condition is satisfied when quarks are light (such as $u$, $d$, or $s$ quarks), or when the transverse momentum ($P_T$) of the jet is much larger than the quark mass.
Thus, if the $P_T$ of the jets is large enough compared to the $b$-quark mass, only gluon jets will be produced exclusively. 
This ``suppression'' mechanism can be used to extract an improved upper limit on the exclusive dijet cross section.
Figure~\ref{fig:exclusive_dpe} (left)
illustrates the method that can be used to determine the heavy-flavor composition of the final sample.

The measured ratio, $D$, of $b$-tagged jets divided by all inclusive jet events is presented in 
Figure~\ref{fig:exclusive_dpe} (right)
as a function of the dijet mass fraction.
A decreasing trend is observed in the data in the large mass fraction region ($R_{jj}>0.7$), 
which may be an indication that the inclusive distribution contains an
exclusive dijet production component.
The fraction of $b$-tagged jets to inclusive jets,
$S=D^{>0.7}/D^{<0.4}$, is measured to be $0.59\pm0.33({\rm stat})\pm0.23({\rm syst})$.
Only four events are present in the last bin and a definite conclusion about exclusive production 
cannot be drawn due to the large statistical and systematical uncertainties.

\vspace*{-0.2cm}

\section{Conclusions}

\vspace*{-0.2cm}

New results from Run~II at Tevatron have been presented.
Exclusive production of dijet events has not yet been found in the data and stringent cross section limits have been set.
A method of extracting the exclusive dijet production from inclusive data using bottom quark jets is presented, exploiting the 
quark/gluon composition of dijet final states.
Given the small number of events collected so far, a definite conclusion cannot be drawn on exclusive dijet production. 
More data to be collected at the Tevatron will help shed light on the exclusive process mechanism and 
allow predictions for future experiments at the Large Hadron Collider.


\begin{thebibliography}{99}
\bibitem{xbj} 
The $x-$Bjorken scaling variable of the struck parton in the antiproton is
$x_{\bar p}=\beta\cdot\xi_{\bar p}$,
where $\beta$ is the pomeron momentum fraction carried by the parton, and $\xi_{\bar p}$ the fractional momentum loss of the scattered antiproton.
For each event, $x_{\bar p}$
is evaluated from the $E_T$ and $\eta$ of the jets using the equation
$x_{\bar p}=\frac{1}{\sqrt s} \sum_{i=1}^nE_T^ie^{-\eta^i}$.
\bibitem{nlo} Sources of background include next-to-leading order (NLO) corrections,
like the process $gg\rightarrow q\bar{q} g$ 
%becomes dominant, 
where
the additional gluon may have large $p_T$, thus making $R_{jj}$ smaller.
\end{thebibliography}
\end{document}